\documentclass[12pt]{iopart}
\usepackage{graphics}
\usepackage{color}
\usepackage{graphicx}
\usepackage{dcolumn} 
\usepackage{bm}
\usepackage{epsfig}
\usepackage{float}
\begin{document}
\title{Pressure-Driven Magnetic Moment Collapse in the Ground State of MnO}
\author{Deepa Kasinathan,$^{1,3}$ 
  K. Koepernik,$^{2,3}$ and W. E. Pickett$^1$}
\address{$^1$Department of Physics, University of California Davis,
  Davis, CA 95616}
\address{$^2$IFW Dresden, P.O. Box 270116, D-01171 Dresden, Germany}
\address{$^3$Max-Planck-Institut f\"ur Chemische Physik fester
Stoffe Dresden, Germany }
\ead{wepickett@ucdavis.edu}
\date{\today}
\begin{abstract}
The zero temperature Mott transition region
in antiferromagnetic, spin S=5/2 MnO is probed using the
correlated band theory LSDA+U method.  The first transition encountered
is an insulator-insulator volume collapse within the rocksalt structure that
is characterized by an unexpected Hund's rule violating `spin-flip' 
moment collapse.  
This spin-flip to S=1/2 takes fullest advantage of the anisotropy of the
Coulomb repulsion, allowing gain in the 
kinetic energy (which increases with
decreasing volume) while retaining a sizable amount of the magnetic
exchange energy.

While transition pressures vary with the interaction strength, the spin-flip
state is robust over a range of interaction strengths and for both
B1 and B8 structures.
\end{abstract}
\submitto{\NJP}
\maketitle

The insulator-metal transition (IMT) in correlated systems (the Mott transition)
is one of the most actively studied topics in condensed matter systems.\cite{RMP}
The original, and most studied, model is that of the single-band Hubbard model (1HM),
characterized simply by bandwidth $W$ and on-site repulsion strength $U$.  Roughly
speaking, 
for $U/W > 1$ it is an insulator characterized by localized
states and local moments, while
for $U/W < 1$ the system is a nonmagnetic metal characterized by 
itinerant states.
In a more general model, the onset of itineracy would lead to increased bonding,
hence a volume collapse at the transition.  It has recently been emphasized that
degenerate multiorbital atoms ($N$ orbitals) with multielectron magnetic moments 
behave very differently.  The critical interaction/bandwidth becomes $(U/W)_c
\approx \sqrt{N}$ or even larger,\cite{koch} due to the increase in conduction
(hopping) channels.  Another issue is that of a possible orbital selective
Mott transition, where only some of the orbitals undergo an IMT 
transition\cite{liebsch,fang,koga} depending on the interactions 
and the anisotropy of
the hopping processes. 

A much less studied question, one we address here, is how a multielectron
local moment disintegrates under reduction of volume. 
While the moment may be considered to be enforced by the
strong interaction $U$ (as in the 1HM), the inter-orbital Hund's coupling is also
a strong factor because it will tend to promote a moment even in the itinerant
phase.  Anisotropic bonding (hopping processes) causes variation in bandwidths, and
the moment collapse may be orbital selective: some but not all
orbitals may become spin-paired (doubly occupied), or selected spins may simply
flip as the kinetic energy overcomes the Hund's coupling but not the Coulomb
repulsion.  Pressure-driven collapses of magnetic signals reported
in $M^{2+}$I$_2$ compounds ($M$=V, Mn, Fe, Co, Ni)\cite{moshe}  and
FeO\cite{moshe2}, initially interpreted as Mott transitions, have been 
reinterpreted as magnetic (dis)ordering transitions rather
than magnetic collapse.\cite{badro} 
In this paper we provide predictions for the moment collapse 
transition in 
anti-ferromagnetic (AFM) MnO 
at T=0, which proceeds by a different route than
any yet envisioned.

The Mott transition in MnO at room temperature has recently been revealed through
transport\cite{PattersonMnO} and spectroscopic\cite{YooMnO,rueff} 
data at high pressure.  
Occurring entirely within the (spin) disordered phase,
there is an insulator-insulator structural transformation 
B1 (rocksalt) $\rightarrow$ B8 (NiAs)
at 90 GPa, followed by an IMT + moment collapse transition at 105 GPa. 
In fact, this (room temperature) Mott transition at 105 GPa is the
first observed for a $3d$ monoxide.  FeO is reported to remain a magnetic insulator
to 143 GPa.\cite{badro}
Unlike for these room temperature experiments where the moments are disordered, 
the magnetic ground state phases we address will be ordered, and the 
resulting symmetry
lowering\cite{gramsch} and reduced fluctuations are found to affect 
the character, and probably
the mechanism, of the transition.

Previous theoretical work on MnO at reduced volume has been carried
out almost entirely in the local spin density approximation 
(LSDA) and generalized
gradient approximation (GGA).\cite{cohen,fang0,fang1,gramsch}   
In these approximations, the small gap at
ambient volume rapidly closes leading to metallization at much too small
a volume.  The resulting occupation of minority $t_{2g}$ states at the
expense of majority $e_g$ states leads to a continuous decrease of the
calculated moment well before the volume collapse, or moment collapse,
transition.  Restricted to the B1 structure, GGA gives a metal-to-metal 
moment collapse from 3.5 $\mu_B$ to 1.2 $\mu_B$ at 150 GPa.\cite{cohen}
The main high pressure phase is expected to be the B8 (NiAs) 
structure,\cite{fang0} and both the crystal and site symmetry and 
structural relaxation have been shown to affect predictions 
strongly.\cite{fang1,gramsch}  Fang {\it et al.} did apply the LSDA+U
method to the high volume phase to improve their picture of the Mott
transition.

To study the pressure behavior of MnO, we have carried out
total energy LSDA+U calculations (described below)
for both low and high pressure
phases. It has recently been shown that metal-to-insulator transitions,
and even charge disproportion and ordering can be modeled realistically
with the LSDA+U method.\cite{kwlee}  
The low pressure structure of MnO is well established: 
it is an antiferromagnetic (AFM) NaCl 
structure with aligned spins in $\langle$111$\rangle$ Mn layers, antiparallel
with adjacent layers (AFMII). In our
calculations, we have neglected the small rhombohedral distortion
angle. 
There are
two simple arrangements of the Mn spins in the B8 phase, ferromagnetic (FM)
or AFM.  Calculations, including structural optimization,
show the AFM phase to
be energetically favorable by 0.2 eV per
formula unit, for
a wide range of pressures.

\begin{figure}[tb]
\begin{center}\includegraphics[%
  clip,
  width=8.5cm,
  angle=0]{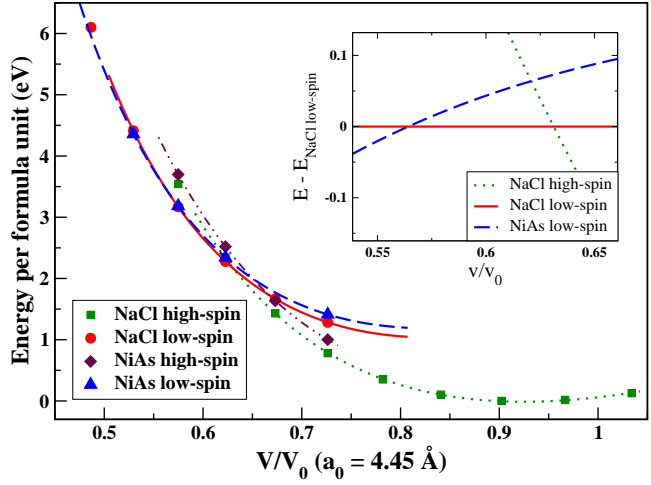}\end{center}
\caption{\label{b1b8ene}The calculated total energy/MnO versus volume for 
the low pressure (NaCl)
and high pressure (NiAs) structures of MnO. The filled symbols denote the 
calculated energies and
the continuous lines are the least square fitted curves to the 
Murnaghan equation of state
for high and low spin configurations respectively. The inset clearly elucidates 
the order of the transitions, NaCl (high-spin) $\rightarrow$ NaCl (low-spin)
$\rightarrow$ NiAs (low-spin).}
\end{figure}

Results we present below use the LSDA+U method\cite{aza}
in the rotationally invariant form\cite{laz} as implemented
in version 5.20 of the full-potential local orbital band
structure method (FPLO\cite{fplo1,fplo2}).
\footnote{
A single numerical basis set for the core states (Mn $1s2s2p$
and O $1s$) and a double
numerical basis set for the valence sector including two $4s$ and $3d$
radial functions, and
one $4p$ radial function, for Mn, and two $2s$ and $2p$ radial functions,
and one $3d$
radial function, for O was used. The semi core states (Mn $3s3p$) are treated as
valence states with a single numerical radial function per
$nl$-shell. We have used the strong local moment form of the LSDA 
double-counting correction\cite{aza,via,solovyev,czyzyk} that has been 
become known as the `atomic-limit' form.\cite{czyzyk}  The
Slater parameters were chosen according to $U=F_0=5.5~\mathrm{eV}$,
$J=\frac{1}{14}(F_2+F_4)=1~\mathrm{eV}$ and $F_2/F_4=8/5$.}).
The implementation of the LSDA+U method in this code has been provided
in detail by Eschrig {\it et al.}\cite{EschrigLSDA+U}
The all-electron aspect of this code is important, since even small-core
pseudopotentials cannot reproduce all-electron results under volume 
reduction.\cite{mitas}  The full-potential aspect can be important also,
on the oxygen site as well as on the Mn site, and the non-spherical 
aspect of the potential will grow as the volume is reduced.
Due to the unexpected nature of the reduced spin state, several results were
checked, and reproduced, using the Wien2k code.\cite{wien2k}
LSDA+U gives two distinct spin states for both B1 and B8 structures:
a low pressure high-spin (HS)
S = $\frac{5}{2}$ configuration and a high pressure, low spin (LS)
S = $\frac{1}{2}$ state.
The equation of state (EOS)
curves are displayed
in Fig.\ref{b1b8ene}.  From the enthalpies
we obtain a first-order magnetic transition from HS-B1 to
LS-B1 at $P_{c1}$=123 GPa, followed by a
structural transition to LS-B8 at $P_{c2}$=130 GPa.
The isostructural volume collapse at $P_{c1}$ is 
$\approx 5\%$. However, these results vary with the choice of $U$ and $J$ (we use 5.5 eV and 1.0 eV
respectively), which is discussed later.
This part of our results has 
been compared with those of other correlated band theory results 
recently.\cite{deepa} 
An unusual feature of the present results is the persistence
of the LSDA+U bandgap up to higher pressures, beyond the observed IMT
at room temperature.\cite{PattersonMnO}  Thus the
magnetic and structural transitions we discuss are always insulator-to-insulator,
and for the magnetically ordered state at T=0.

\begin{figure}[tb]
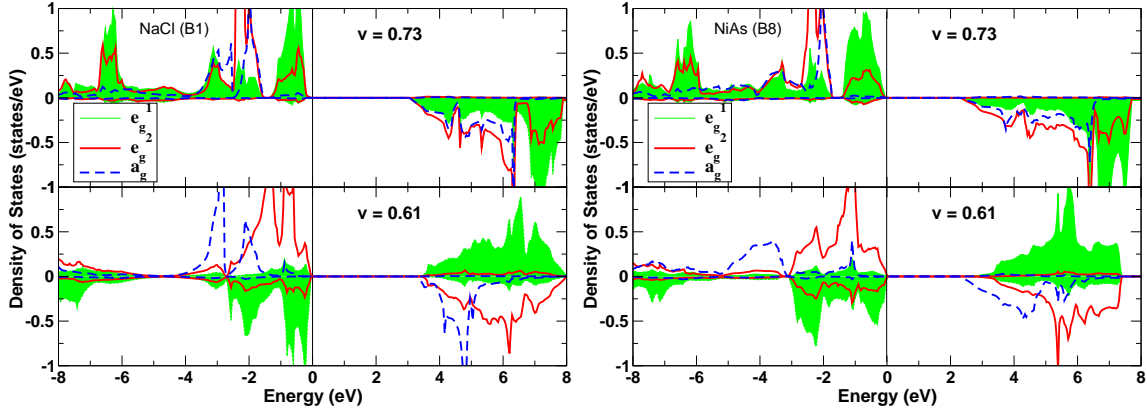

\begin{center}
\includegraphics[clip,width=7.5cm,
  angle=-0]{Fig2a.eps}
\includegraphics[clip,width=7.5cm,
  angle=-0]{Fig2b.eps}
\end{center}
\caption{\label{ldaudos}
LSDA+U DOS, at the indicated volumes,
projected onto symmetrized Mn 3$d$ orbitals
in (left panels) the rhombohedral B1 AFMII
phase and (right panels) the B8 AFM structure.  In each case, the
top subpanel is for the high spin state, while the bottom is for the
low spin state.  See the text for the definition of the $a_g, e_g^1,
e_g^2$ labels.
The overriding feature is the spin-reversal of the $m=\pm 1$
$e^{1}_{g}$ orbitals between the two volumes.  Broadening of the
$a_{g}$ states in the LS-B8 DOS is due to the direct Mn-Mn $d$-overlap
in the $z$ direction in B8 structure (Mn lies on a simple hexagonal
sublattice).
}
\end{figure}

To help in understanding the mechanism of the transition, 
the densities of states (DOS) projected onto each of the
$\ell=2$ irreducible representations are displayed in Fig. \ref{ldaudos}, 
referenced to the rhombohedral axis of the B1 AFMII phase, and equivalently
the hexagonal axis of the B8 structure.  Not evident from this figure are
two potentially important features.\cite{kunes} First,
the `charge transfer' energy increases
as the volume decreases; that is, the O $2p$ levels drop in energy 
relative to the Mn $3d$ states, reflecting an {\it increased} tendency
toward the fully ionic limit that competes with the increased 
hybridization as the Mn and O ions approach each other.  Secondly,
the crystal field splitting between $E_g$ and $T_{2g}$ states increases
under pressure, which competes with Hund's exchange and with correlation
effects.

In each case the
Mn site symmetry splits the 5 $d$-orbitals into two doublets 
$e^{|m|}_{g}$, $|m|$=1, 2, with $e^{1}_{g} \rightarrow \{xz, yz\};
e^{2}_{g} \rightarrow \{x^{2}-y^{2}, xy\}$,
and a singlet
$a_{g}\rightarrow 3z^{2} - r^{2}~(m=0)$. 
The actual state realized
at a given volume or structure is characterized by two $e_{g}$ pairs,
obtained by unitary mixing, schematically:
\begin{equation}
\eqalign{e_g^a = \cos\beta~e_g^1 - \sin\beta~e_g^2, \cr
e_g^b = \sin\beta~e_g^1 + \cos\beta~e_g^2. }
\end{equation}
We have observed and quantified the mixing angle $\beta$ versus pressure,
but near the critical pressure it simplifies to $\beta \approx 0.$ 

\begin{figure}[tb]
\begin{center}
\includegraphics[clip,width=6.5cm, 
  angle=-0]{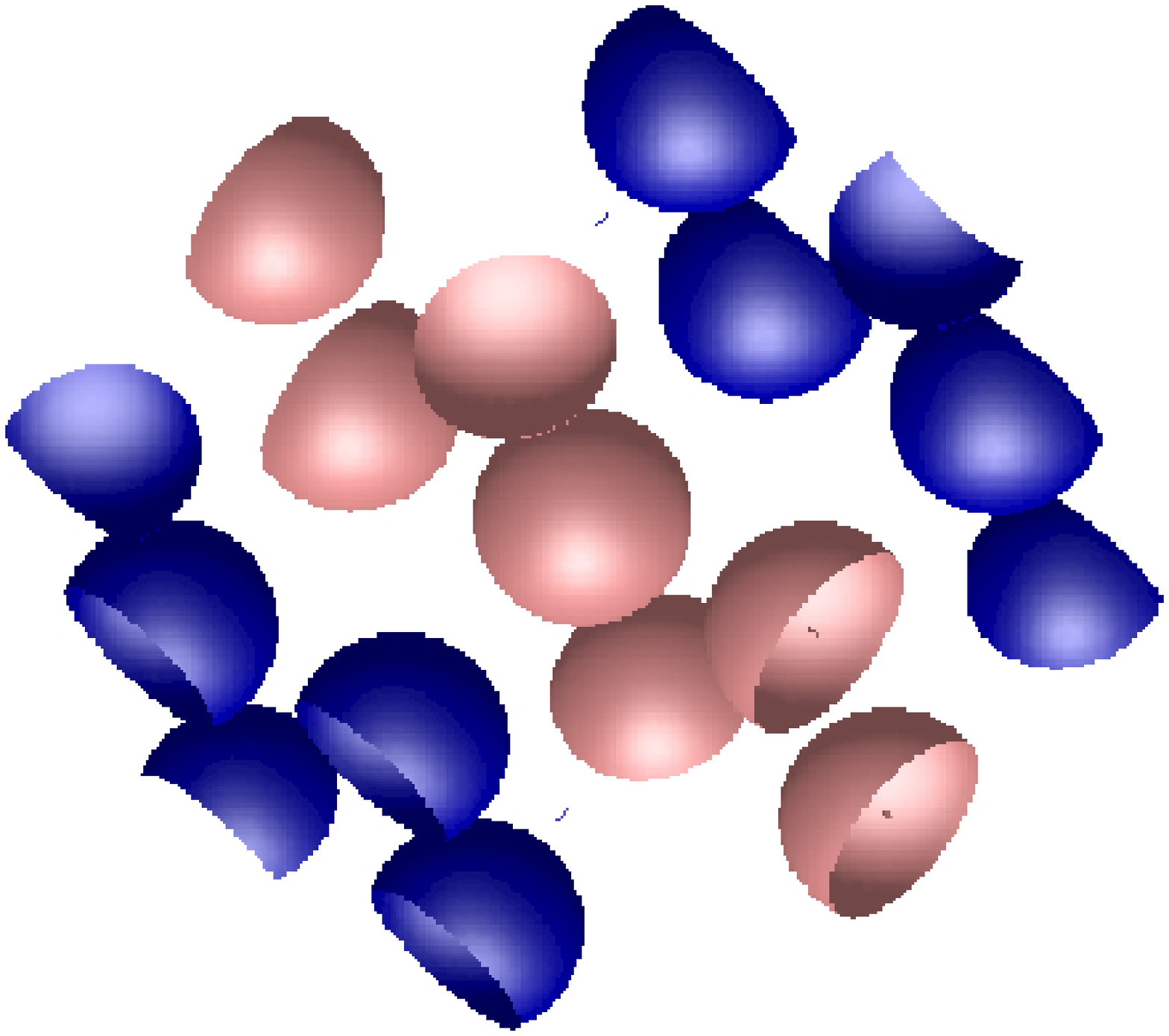}
\includegraphics[clip,width=7.0cm,
  angle=-0]{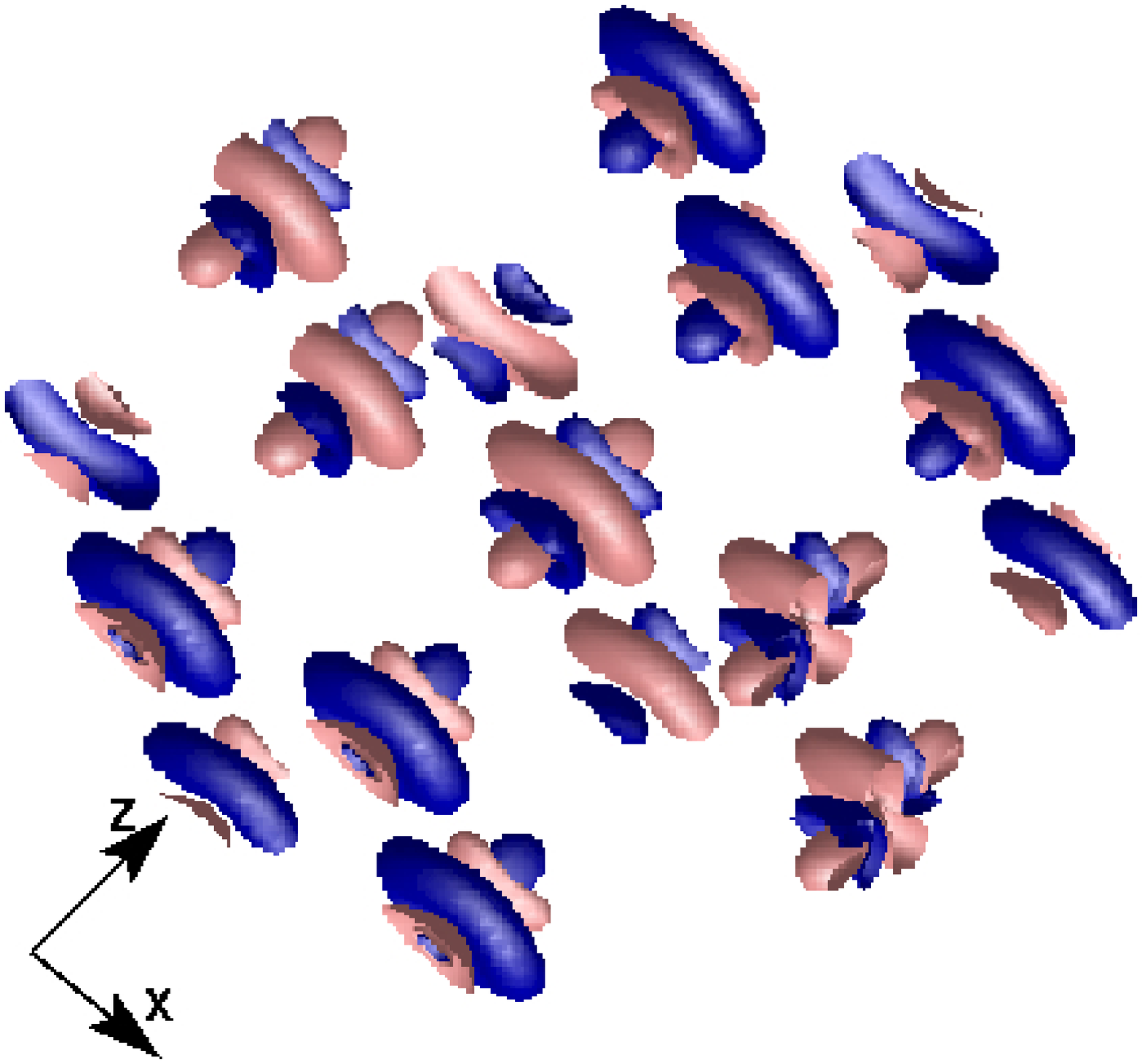}\end{center}
\caption{\label{spindens} 
Isosurface plots of Mn ion spin density, with red and blue shading
indicating opposite sign. Left: before collapse, showing
the spherical S=5/2 ion, with (111) layers of aligned spins. Right:
after collapse, revealing the anisotropic S=1/2 ion.  The magnetic order 
remains AFMII.}
\end{figure}

The HS states in both B1 and B8 structures are simple
-- each 3$d$ orbital is filled once with  
spins aligned leading 
to an S=5/2 spherical ion. 
The LS DOS for both structures reveal the
essence of the HS-LS transition: the LS state is obtained
by simply flipping the spins of the $e^{1}_{g}$ orbitals. This result shows how
the LSDA+U method differs in an essential way from LSDA, where the moment decreases 
continually with volume,\cite{cohen} metallization occurs at low 
pressure, and decrease of
the moment implies rapid collapse of the exchange splitting, resulting in 
doubly occupied orbitals with zero net spin.
In this LS state, each of the 3$d$ orbitals remains singly
occupied, the charge density remains spherical while the spin density 
becomes highly anisotropic, as illustrated vividly in Fig.
\ref{spindens}.  

To understand the origin of this spin-flip state we have performed
an analysis of the LSDA+U method. The flavor we have used is the
``atomic limit'' (AL).\cite{via,solovyev,czyzyk}
We consider first the pressure-induced change in kinetic energy relative to 
the potential energy of the HS and LS states.  We note first that the
$e_g^1$ states, whose spins flip in the LS state, are $\frac{2}{3}E_g$
and only $\frac{1}{3}T_{2g}$ in terms of the cubic states we are more
familiar with.  For the $e_g^2$ pair this ratio is opposite, and close
to the overall mean ($\frac{2}{5}E_g, \frac{3}{5}T_{2g})$.  It is
the cubic $E_g$ states (and consequently the $e_g^1$ pair) that have
the strongest ($dp\sigma$) overlap with O ions, and thus are 
most affected by pressure and give the greatest gain in kinetic energy.  

Now we consider the effects of the very large anisotropy of the LS Mn ion.
To separate the effects of $U$ from those of $J$, we split
the energy expression into the isotropic interaction, 
and the remaining
anisotropic part: $E^{AL} = E^{aniso} + E^{iso}$.  
The isotropic part reads
\begin{equation}
E^{iso}= \frac{1}{2}\left(U-J\right)\sum_{s}\mathrm{Tr}
   \left[n^{s}\left(1-n^s\right)\right] \geq 0,
\end{equation}
where $n^{s}$ is the spin-dependent occupation number
matrix of the $3d$ shell, and Tr denotes a trace of the orbital indices.
It is easily seen that $E^{iso}$ takes
its minima for integer occupations, in which case we get $E^{iso} = 0$.
However, the resulting potential matrix, to be added to the 
Kohn-Sham Hamiltonian,
is non-zero and has the effect of lowering occupied orbitals by 
$-\frac{1}{2}\left(U-J\right)$,
while raising unoccupied orbitals by $\frac{1}{2}\left(U-J\right)$.
This separation stabilizes insulating magnetic solutions, and  also
favors HS exchange energy
contributions from the magnetic part of the LSDA functional. 
This tendency opposes
the observed HS-LS transition, but operates equally independent of volume.
The action of the isotropic term is often the dominating effect
of the LSDA+U method, but the common practice of discussing the effects
of the LSDA+U method in the isotropic limit misses the physics of this transition,
as we now illustrate.

The anisotropic term reads (in the representation in which $n^s$
is diagonal, for simplicity) 
\begin{eqnarray}\label{eaniso}
E^{aniso}=\frac{1}{2}\sum_{ss^{\prime}}\sum_{\mu m}n_{m}^{s}
  \left[\Delta U_{m\mu}-\delta_{ss^{\prime}}\Delta J_{m\mu}\right]
  n_{\mu}^{s^{\prime}}.
\end{eqnarray}
The interaction
matrix elements are defined as $\Delta U_{m\mu}=w_{m\mu}^{m\mu}-U$,
$\Delta J_{m\mu}=w_{m\mu}^{\mu m}-J-\left(U-J\right)\delta_{m\mu}$ in
terms of the full matrix interaction $w_{m\mu}^{m\mu}$, 
and these differences do {\it not} contain $U$ 
(whose effect is included entirely in the isotropic term).
These differences describe {\it pure anisotropy}; summation of either index of either one
gives a vanishing result. Thus a filled spin subshell will not 
contribute to $E^{aniso}$, as expected intuitively.

The HS state is favored by the 
LSDA spin polarization.  However, the LS state still has all fully
polarized orbitals, so the energy difference will be much smaller than
for usual LSDA S=5/2 and S=1/2 moments. 
The resulting ratio of exchange energies can be estimated
from the integral over the square of the spin density, which gives a
value $E_{1/2}/E_{5/2} \approx 0.31$. This reduction of exchange energy in
the LS state is much less dramatic than the estimate from the simple
formula $E_x = -\frac{J_{\mathrm{Stoner}}}{4} M^2$. A more detailed
explanation of this energy ratio is given in the Appendix.

The LS configuration
requires the $a_{g}$ orbital to be singly occupied.
The remaining
$4$ electrons then are distributed in pairs among the $e_{g}^{a,b}$ doublets.
Analysis shows that the  
two occupation patterns for which the electrons doubly occupy either
the $e_{g}^{1}$ or $e_{g}^{2}$ orbitals have the same anisotropy
energy of about $E^{aniso} \geq -0.3J$; 
a spread of energies arises from allowed mixing of $e_{g}$ symmetries.
The resulting spin density of such states has $a_{g}$-derived shape.
 
The remaining two patterns are
obtained by occupying $e_{g}^{a}$ in the up-channel and $e_{g}^{b}$
in the down-channel or vice versa, both of which results in strongly 
anisotropic spin densities. The lowest energy of $E^{aniso} \approx -1.85J$
is obtained when the $e_{g}^{2}$ and $a_{g}$ are occupied in the
same spin channel, while $e_{g}^{1}$ is occupied in the opposite channel
(note, the mixing angle $\beta$=0).
The other solution ($e_{g}^{1\uparrow}\parallel a_{g}^{\uparrow}$)
has $E^{aniso} \approx -1.14J$. The dependence of the energies on the
mixing angle complicates the discussion. 

However, there
is a gap of $\approx 0.84J$ between the spin-flipped and non spin-flipped
solutions, which is not closed by any mixing. It turns out that this gap
is due to the density-density anisotropy ($\Delta U$ in
Eq. (\ref{eaniso})), which pushes the (density-) non-spherical non spin-flipped
patterns  up in energy, while it is zero for the (density-) spherical
spin-flipped patterns. The exchange and self-interaction-correction
 contributions to
the anisotropy ($\Delta J$ in Eq. (\ref{eaniso})) is nearly of the same size for
the non spin-flipped and spin-flipped occupation patterns, and hence not
changing the energy separation of these pattern classes. 
However, it further discriminates the two spin-flipped patterns.

The anisotropy (which is solely controlled by, and proportional to, $J$ in the LSDA+U
method) of the interaction favors
an occupation pattern which maximizes the spatial distance between
the electrons (under the constraint of $S = \frac{1}{2}$), while the
isotropic term $\propto U-J$ merely selects insulating over metallic solutions.
We indeed find that both the spin-flipped and non-spin
flipped solutions may be found in LSDA+U calculations, separated by
an energy of the order derived here.  When $J$ is decreased to zero, 
the energy difference between these
solutions shrinks, leaving only the 
LSDA anisotropy energy difference, which turns out to be
very small: about $0.3$ eV at ambient pressure and
decreasing to zero close to the transition. Of course the LSDA part of the
functional includes some anisotropy effects, however the main difference in the
influence of the LSDA-anisotropy and the $J$-anisotropy
Eq. (\ref{eaniso}) is its action on the Kohn-Sham states. 
The LSDA potential contributions act on all states, while the LSDA+U
potential  matrix acts orbital selective. It is this very selectivity, which
makes  the whole LSDA+U machinery work, by mimicking the suppression of
occupation number fluctuations due to correlations. 
In the same way as the Hubbard band split is not
attainable in LSDA, the anisotropy effects are largely suppressed. 
This suppression is nicely confirmed by the observation of the vanishing
energy difference between the two LS solutions as described above.

With increasing pressure the kinetic energy gain becomes more and more
competitive with the exchange energy due more to the increasing crystal
field splitting than to the bandwidth. This competition usually leads to
a (partial) collapse of the magnetic moment. LSDA calculations give a moment of
about 1.5 $\mu_B$ at our transition pressure, which is far from the HS
value of 5 $\mu_B$ thus clearly showing that at the transition pressure
first Hund's rule is strongly suppressed. In the LSDA+U method, the
isotropic term forces the HS solution to have full spin-moment, while
the LS solution allows a larger gain in kinetic energy, hence bringing
along a transition from HS to LS at some pressure. 
The anisotropy contribution of LSDA+U is zero for the HS state and
negative for the flipped LS state. This will further lower the energy of the
flipped LS state against the HS state, resulting in a lower transition
pressure. Moreover, this anisotropy contribution is smaller for a
non spin-flipped solution, which rules out this solution.
Since the anisotropy term offers a way to keep a sizable amount of
magnetic exchange energy, while gaining kinetic energy, it is this
``unusual'' state, which is realized after the transition.

We have described here how LSDA+U energies for MnO under pressure 
predict an unexpected mode of collapse of the
Mn moment at zero temperature: each of the $3d$ orbitals remains polarized, and
an S=5/2 to S=1/2 reduction arises from a simple spin-flip
of the symmetry-determined $e_g^1$ doublet that has the strongest overlap with
neighboring O $2p$ orbitals.  
This S=1/2 moment in the high pressure
phase is consistent with the interpretation of xray emission data by
Rueff {\it et al.}\cite{rueff}; Yoo {\it et al.} were less specific
about the value of the (clearly small) high pressure moment but
presumed total collapse.
The partial spin-flip collapse obtained here occurs
in both the B1 (rocksalt) and B8 (NiAs) structures, calculated by two
different codes, and occurs at
similar volumes. 

\begin{figure}[tb]
\begin{center}
\includegraphics[clip,width=7.5cm, 
  angle=-0]{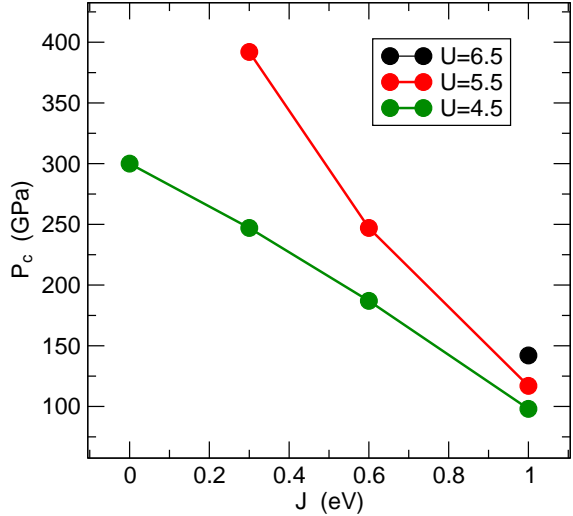}\end{center}
\caption{\label{PcUJ} 
Dependence of the high-spin to `spin-flip' low-spin transition on the parameters, $U$
and $J$.}
\end{figure}

The transition we find is first-order and insulator-to-insulator,
both of which insinuate the smallness
of fluctuation effects and make the LSDA+U approach an appropriate one.
Due to the imprecisely known values of $U$ and $J$ there is an
associated uncertainty for the calculated equation of state, which depends on
the values chosen.
The functional dependence of these interaction energies ($U$, $J$) on the
density is not known and we have neglected the volume dependence.
The Wien2k code\cite{wien2k} includes a constrained LSDA algorithm that
enables calculation of $U$, resulting in a value of 6.5 eV for
both for the 
equilibrium volume and close to the transition.  
To identify the range
of variation,
we performed EOS calculations for various values of $U$ and $J$.
The resulting transition pressures are depicted in Fig. \ref{PcUJ}.
As should be expected from the importance of the correlation corrections, the transition
pressure $P_c(U,J)$ is quite dependent on the parameters. For all values of $U$,
$P_c$ decreases with increasing $J$. The strong variation confirms our analysis
above, which identified the anisotropy (proportional to $J$) as crucial in
determining the ground state at a given volume.
The energy 
separation of the HS and LS curves decreases with $J$, and this change
decreases the transition pressure (the shape dependence of the
energy curves on $J$ is minor).  On the other hand, the potential matrix element effects
(which shift the corresponding eigenvalues) tend to increase the Hubbard splitting with
increasing $U$, which moves to stabilize the HS solution against the LS solution 
and leads to a monotonic increase of $P_c$ with $U$.  Note that for reasonable
values of $J$=0.6--1 eV, the dependence on $U$ lessens compared to 
stronger variation for unreasonably
small values of $J$. The very much to large transition pressure for
$J=0$ eV compared to experiment is a strong argument that the anisotropy
effects are not sufficiently described by the LSDA part of the
functional and hence have to be included in an orbital selective manner.

Although the occupation number fluctuations at this transition should not
be a big factor, as the volume is reduced and the bandwidth increases, these
fluctuations will tend to increase.  The effect can be
modeled by adopting a smaller $U$ than the ambient pressure value; hence
the value of 5.5 eV that was used for most results presented here becomes
justified.
On the other hand, the value of $J$ is only weakly screened by the
environment, and it
is sensible to chose a volume-independent value that resembles the atomic/ionic
situation, where one usually finds $J$=0.7--1 eV for transition
metals.  Altogether, the calculated transition pressures vary between
100 and 170 GPa for a reasonable choice of $J$, which is quite
satisfactory given the uncertainties of the LSDA+U approach.  One should
also keep in mind that the zero temperature transition is expected to
occur at a higher pressure than at room temperature.

However, we want to stress our main point. Although the LSDA+U method
is not capable of precise predictions of $P_c$ due to the uncertainties
just discussed, the {\it spin-flip character}
of the predicted ground state is highly stable against changes of the
interaction parameters. For all parameter sets corresponding to the data
points in Fig.  \ref{PcUJ} we obtain the spin-flip LS solution as the ground
state. 

Now we summarize.  For MnO at T=0 in the AFM ordered phase, an unusual
moment collapse $S = 5/2 \rightarrow 1/2$ is predicted before the Mott 
transition (metallization, or itineracy of the $3d$ states).  
These results are robust: the spin-flip state is obtained by two different
codes, and for a substantial range of choices of $U$ and $J$.  The value of 
$U=5.5$ eV used here corresponds to $U/W \sim$1.5 in terms of the full $3d$ bandwidth $W$.  
The role of $J$ is
central to this transition, but in an unexpected way.  Hund's first rule,
which is encouraged by the spin-exchange aspect of $J$, 
is violated at the transition, whereas the
anisotropic Coulomb repulsion that is proportional to $J$ becomes
the driving force.  Together with an orbitally-dependent increase in kinetic energy,
the result is an orbitally-selective spin-flip collapse of the moment at
an insulator-to-insulator transition.  

The order and type of transitions under pressure we obtain differ from that
observed at room temperature.\cite{PattersonMnO,YooMnO}  
It is established however that 
structural phase boundaries can be strongly temperature dependent in transition
metal oxides,\cite{mao} so there is no contradiction.  
The predicted pressure range
is accessible to diamond anvil cell experiments, and the ordered-phase moment
can be probed by M\"ossbauer spectroscopy.\cite{moshe}

\ack
We acknowledge stimulating interaction on this topic with J. Kune\v{s}, 
B. Maddox, A. K. McMahan, R. T. Scalettar, E. R. Ylvisaker, and C. S. Yoo.
Work at UCD was supported by Department of Energy
grant DE-FG03-01ER45876.  This collaboration was stimulated by DOE's
Computational Materials Science Network, and we acknowledge important 
interactions within
the Department of Energy's
Stewardship Science Academic Alliances Program. This work was further
supported by the Emmy Noether program.

\appendix
\section*{Appendix}
\setcounter{section}{1} 

The modern flavour of LSDA+U explicitly excludes first Hund's rule
from the expression added to the functional, arguing that this
contribution is dealt with better within the LSDA part of the
functional.
In order to obtain estimates of this LSDA-contribution, we will derive
an approximate expression for the LSDA xc-energy $E_{xc}$ suitable for the
situation we discussed in this paper.

We expand $E_{xc}$ up to second order in variations of the density.
A natural choice for a reference density would be a spherically averaged
non-spin-polarized density around the atom center. We denote the reference
density by $\rho_0$. Our interest is in the effect of different orbital
occupations on the magnetic ion. We can describe the spin-density of the
$l$-shell by
\begin{equation*}
  \rho^{s}(\mathbf{r}) = \sum_{mn} \phi_{lm}(\mathbf{r}) n^{s}_{m,n}
  \phi_{ln}^*(\mathbf{r})  \;,
\end{equation*}
where $n^{s}_{mn,}$ denotes the generalized occupation number matrix for spin
$s=\pm 1$, $\phi_{lm}(\mathbf{r})$ are suitable orbitals and the indices
$m,n$ run over the orbitals of the shell. In the usual manner one can write the
orbitals as fixed atom-like functions
\begin{equation*}
  \phi_{lm}(\mathbf{r})=R_{l}(r) Y_{lm}(\hat{r}) \; ,
\end{equation*}
putting the flexibility into the occupation number matrix.
We can introduce the occupation number matrix $\underbar{n}=\sum_{s}
\underbar{n}^{s}$ for the charge density $\rho(\mathbf{r})=\sum_s \rho^s(\mathbf{r})$ and
the occupation number matrix $\underbar{m}=\sum_{s} s \underbar{n}^{s}$
for the magnetization density $m(\mathbf{r})=\sum_s s
\rho^s(\mathbf{r})$. The particle number of the spin channel $s$ is 
$N^{s} =\Tr \underbar{n}^s$, which gives the  total number of particles
as $N= \sum_{s} N^{s}$ and the magnetic moment as $M= \sum_{s} s N^{s}$.
The spherical and spin average of $\rho^s$ is included in
the definition of the reference density, hence the density variation due
to different occupation patterns is
\begin{equation}\label{eq:dens_variation}
 \delta \rho^{s}(\mathbf{r}) = [R(r)]^2 \sum_{mn} Y_{lm}(\hat{r}) \delta
 n^{s}_{m,n} Y_{ln}^*(\hat{r}) \; ,
\end{equation}
where the variation of the occupation numbers $\delta \underbar{n}^s$ is
measured with respect to  the averaged occupation numbers ($\Delta=2l+1$)
\begin{equation*}
  n^{s}_{mn,0}=\delta_{mn} \frac{N}{2 \Delta}\;.
\end{equation*}
Since the reference density is non-polarized the variation of the
magnetization density equals the magnetization density itself: 
 $\underbar{m}=\delta \underbar{m}$, $m(\mathbf{r})=\delta m(\mathbf{r})$.

To keep things simple, we restrict the discussion to the local-density
form of the xc-energy
\begin{equation*}
  E_{xc}=\int \rho \varepsilon(\rho^{+},\rho^{-}) d^3r\;.
\end{equation*}
The second order variational expansion around the reference density then reads 
\begin{eqnarray}
  E_{xc}&=&E_{xc,0} 
+ \int V_{xc,0}(r) \delta \rho(\mathbf{r})  d^3r
+ \int B_{xc,0}(r)  m(\mathbf{r})  d^3r \nonumber\\
&& \!\!\!\!\!\! \!\!\!\!\!\! + \frac{1}{2} \int \left [ P_{0}(r) (\delta \rho(\mathbf{r}))^2 
 +  2 Q_{0}(r) (\delta \rho(\mathbf{r}) m(\mathbf{r}))
+ K_{0}(r) ( m(\mathbf{r}))^2
\right ] d^3r \;. \nonumber
\end{eqnarray}
The xc-potential and the second order xc-kernels are spherical due to
our spherical reference density  ($V_{xc,0}(\mathbf{r})=V_{xc,0}(r)$)
and the xc-field is zero, since the reference density is non-polarized. 
Using this information and the shape of the density variation
Eq. (\ref{eq:dens_variation}) we arrive at
\begin{eqnarray}
  E_{xc}&=&E_{xc,0} + v_0 \delta N \nonumber\\
&& \!\!\!\!\!\! \!\!\!\!\!\!  \!\!\!\!\!\! + \frac{1}{2} \sum_{mnm^{\prime}n^{\prime}}
\left [  
  p_{0} \delta n_{mn} \delta n_{m^{\prime}n^{\prime}}
 +  2 q_{0} \delta n_{mn} m_{m^{\prime}n^{\prime}}
+ k_{0}  m_{mn} m_{m^{\prime}n^{\prime}}
\right ]  a_{mn,m^{\prime}n^{\prime}}\nonumber
\end{eqnarray}
with the variation of the particle number $\delta N= \Tr \delta \underbar{n} $,
with the angular coefficients
\begin{equation}
  a_{mn,m^{\prime}n^{\prime}}=4\pi \int 
Y_{lm}(\hat{r})Y_{ln} (\hat{r})
Y_{lm^{\prime}}(\hat{r})Y_{ln^{\prime}} (\hat{r}) d \Omega
\end{equation}
and with the radial integrals
\begin{equation*}
  v_{0}= \int V_{xc,0}(r) [R_{l}(r)]^2 r^2 dr
\end{equation*}
\begin{equation}
  k_{0}=\frac{1}{4\pi} \int K_{0}(r) [R_{l}(r)]^4 r^2 dr
\end{equation}
\begin{equation}
  p_{0}=\frac{1}{4\pi} \int P_{0}(r) [R_{l}(r)]^4 r^2 dr
\end{equation}
\begin{equation}
  q_{0}=\frac{1}{4\pi} \int Q_{0}(r) [R_{l}(r)]^4 r^2 dr \;.
\end{equation}
For $l \leq 2$ and real spherical harmonics we obtain
\begin{equation}
  a_{mn,m^{\prime}n^{\prime}}=\frac{\Delta}{\Delta+2}
\left [
\delta_{mn} \delta_{m^{\prime}n^{\prime}} 
+2 \delta_{mn^{\prime}} \delta_{nm^{\prime}} 
\right] \;,
\end{equation}
which leads to the simple expression
\begin{eqnarray}
  E_{xc}&=&E_{xc,0} + v_0 \delta N \nonumber\\
&& + \frac{\Delta}{2(\Delta+2)} 
\left [  
  p_{0} (\delta N)^2  +  2 q_{0} \delta N M + k_{0}  M^2  \right . \nonumber\\
&& \left . 2 p_{0} \Tr (\delta \underbar{n})^2
          +4 q_{0} \Tr (\delta \underbar{n} \underbar{m})
          + 2 k_{0} \Tr (\underbar{m})^2)   \nonumber
\right ]  \;.
\end{eqnarray}
In discussing the MnO case, we consider only $3d^5$ occupation patterns,
hence $\delta N=0$. For a spherical  magnetic occupation pattern we
have to set $\delta n^{s}_{mn}=\delta_{mn} \frac{s M}{\Delta}$, $\delta
n_{mn}=0$, $\delta m_{mn}=\delta_{mn} \frac{M}{\Delta}$, resulting in 
$ \delta E_{xc}=  \frac{k_{0}}{2}  M^2$,
which suggest the interpretation $k_0=-\frac{I}{2}$ with the Stoner
parameter $I$.

For the HS and spin-flipped LS pattern (SF-LS) we get $\delta \underbar{n}=0$,
since these patterns correspond to a spherical charge density. The
magnetic occupation numbers are diagonal and equal to $\pm1$, hence
$\Tr \underbar{m}^2=\Delta$. So we get 
 \begin{equation*}
  \delta E_{xc}^{\mathrm{HS}}=  -\frac{I}{4} M^2=  -\frac{I}{4} 25
\end{equation*}

\begin{equation*}
  \delta E_{xc}^{\mathrm{SF-LS}}=  -\frac{I}{4} M^2 -\frac{I}{4}  
\frac{2}{\Delta+2}\left( \Delta \Tr \underbar{m}^2 -M^2\right)
=  -\frac{I}{4} 1 -\frac{I}{4} \frac{48}{7}
 =  -\frac{I}{4} \frac{55}{7} \;.
\end{equation*}
The SF-LS energy contains a large contribution, which is related to the
non-sphericity of the spin-density. It accounts for $\approx 7/8$-th of
the whole xc-energy of this configuration. The ratio between the
energies of these two configurations is
$E_{\mathrm{SF-LS}}/E_{\mathrm{HS}}=11/35 \approx 0.31$. 

For the non spin-flipped pattern (NSF-LS) the diagonal occupation number
matrices read $\delta n =(1,-1,0,-1,1)$ and $\delta m =(0,0,1,0,0)$,
which gives for the change in xc-energy
\begin{eqnarray*}
  \delta E_{xc}^{\mathrm{NSF-LS}} &=& -\frac{I}{4} M^2
 -\frac{I}{4} \frac{2}{\Delta+2} \left ( \Delta \Tr (\underbar{m})^2 -  M^2 \right )  
 -\frac{I}{4} \frac{p_{0}}{k_{0}} \frac{2\Delta}{\Delta+2}  \Tr (\delta \underbar{n})^2
 \nonumber\\
   &=& -\frac{I}{4} 1
 -\frac{I}{4} \frac{8}{7}   
 -\frac{I}{4} \frac{p_{0}}{k_{0}} \frac{40}{7}  
\;.
 \nonumber
\end{eqnarray*}
Again, the second term is due to the non-sphericity of the spin
density. However, it is much smaller than for the flipped case. The
third term, proportional to $p_0$ is related to the non-sphericity of
the charge density of the shell. Estimates of $p_0$ from actual
calculations give a value of $p_0 \approx \frac{1}{2} k_0$, hence the
third term is roughly of the same size as the sum of the first two
terms: $ \delta E_{xc}^{\mathrm{NSF-LS}} \approx  -\frac{I}{4} 5 $,
which is only two-thirds of the energy of the spin flipped case.

\end{document}